\documentclass[prc,aps,amsfonts,superscriptaddress,showpacs,showkeys,floatix]{revtex4-2}

\usepackage{epsfig}
\usepackage{amssymb}
\usepackage[utf8]{inputenc}
\usepackage{graphicx}
\usepackage{dcolumn}
\usepackage{bm}
\newcommand{\be}{\begin{equation}}
\newcommand{\ee}{\end{equation}}
\newcommand{\bea}{\begin{eqnarray}}
\newcommand{\eea}{\end{eqnarray}}
\newcommand{\ba}{\begin{array}}
\newcommand{\ea}{\end{array}}
\newcommand{\bi}{\begin{itemize}}
\newcommand{\ei}{\end{itemize}}
\newcommand{\lan}{\langle}
\newcommand{\ran}{\rangle}

\begin{document}
\title {Nuclear Matrix Elements for Heavy Ion Sequential Double Charge Exchange Reactions}
\author{Horst Lenske}
\affiliation{Institut f\"ur Theoretische Physik, Justus--Liebig--Universit\"at Giessen, Germany}
\email[Correspondence:]{horst.lenske@physik.uni-giessen.de}

\author{Jessica Bellone}
\affiliation{Istituto Nazionale di Fisica Nucleare, Laboratori Nazionali del Sud, I-95123 Catania, Italy}
\author{Maria Colonna}
\affiliation{Istituto Nazionale di Fisica Nucleare, Laboratori Nazionali del Sud, I-95123 Catania, Italy}
\author{Danilo Gambacurta}
\affiliation{Istituto Nazionale di Fisica Nucleare, Laboratori Nazionali del Sud, I-95123 Catania, Italy}
\collaboration{The NUMEN Collaboration, LNS Catania, I-95123 Catania, Italy.}

\begin{abstract}
The theoretical approach to a sequential heavy ion double charge exchange reaction is presented. A brief introduction into the formal theory of second-order nuclear reactions and their application to Double Single Charge Exchange (DSCE) reactions by distorted wave theory is given, thereby completing the theoretical background to our recent work \cite{Bellone:2020lal}. Formally, the DSCE reaction amplitudes are shown to be separable into superpositions of distortion factors, accounting for initial and final state ion--ion interactions, and nuclear matrix elements. A broad space is given to the construction of nuclear DSCE response functions on the basis of polarization propagator theory. The nuclear response tensors resemble the nuclear matrix elements of $2\nu\beta\beta$ decay  in structure but contain in general a considerable more complex multipole and spin structure. The QRPA 
theory is used to derive explicit expressions for nuclear matrix elements (NMEs). The differences between the NME of the first and the second interaction vertexes in a DSCE reaction is elucidated. Reduction schemes for the transition form factors are discussed by investigating the closure approximation and the momentum structure of form factors. DSCE unit strength cross sections are derived.
\end{abstract}
\maketitle

\section{Introduction}\label{sec:Intro}

The study of higher-order nuclear processes is a very demanding field of research, especially when they are driven by hadronic interactions. The~complexities are introduced by the equal importance of nuclear many-body aspects and the effective nature of in-medium low-energy nuclear interactions. Since both sectors are intimately intertwined, a~clear separation of effects is hardly possible, thus inhibiting a straightforward  perturbative approach. Still, over~the years, nuclear reactions and structure theory have developed a tool box of methods allowing now systematic investigations of rare nuclear processes and their spectroscopy. Phenomenological approaches and the meanwhile rather successful many-body approaches based on effective nuclear field theory have reached a level of accuracy that fine details of nuclear spectroscopy are now accessible---and predictable---by theory. Our recent work on a first-time quantitative description of heavy ion double charge exchange (DCE) data on microscopic grounds~\cite{Bellone:2020lal} is a prominent example for that kind of~achievement.

Although there is a general consensus on the importance of studying higher-order nuclear processes, detailed studies are rare. Recent examples are the multi-phonon description of the extremely rare nuclear double-gamma emission~\cite{Soderstrom:2020iaz} and, using similar nuclear structure methods, the~study of the quenching of low-energy Gamow--Teller strength~\cite{Gambacurta:2020dhb}. On~the experimental side of nuclear reactions, the~NUMEN project~\cite{Cappuzzello:2018wek} is the trend-setting case of a research project fully devoted to a higher-order nuclear reaction, namely to investigate nuclear DCE reactions with heavy ion beams,  aiming to make  an independent probe for the nuclear matrix elements of nuclear double beta decay (DBD) available.

Higher-order reactions are of interest for nuclear reaction physics. They have a high potential to reveal rare reaction mechanisms hitherto undiscovered because they are not present or suppressed in first-order processes. Heavy ion second-order reactions have rarely to never been used as spectroscopic tools. An~exception is statistical multi-step reactions in the pre-equilibrium region of nuclear spectra, which have been studied in the past for light and heavy ion reactions using Multi-step Direct Reaction (MSDR) theory~\cite{Tamura:1982zz,Lenske:1983tnq,Lenske:2001rfn}.  In~\cite{Ringbom:1997bjw,Ramstrom:2004fem}, the MSDR scheme was used to study neutron-induced single charge exchange (SCE) reactions in the continuum region of the spectra in a two-step approach. In~the present context, we are interested in a much more selective case, namely on  $A(N,Z)\to A(N\pm 2,Z\mp 2)$ reactions leading to the discrete part of the spectrum. These reactions take place in a complementary manner in the projectile and the target system. Hence, for~a complete description, both nuclear systems must be described and understood simultaneously. From~a general point of view, however, that apparent complication can be considered an advantage because it allows us to probe two DBD processes simultaneously in a single reaction, namely a $2\beta^+$ transition in one nucleus by a complementary $2 \beta^-$ transition in the other nucleus. Thus, DCE reactions pose a double challenge to nuclear~theory.

The reaction mechanism of a DCE reaction with composite nuclei is by no means obvious. In~principle, DCE reactions can proceed either by mutual nucleon transfer processes or by acting twice with the isovector nucleon--nucleon (NN) interaction. A~first detailed discussion on that important issue is found in our recent review~\cite{Lenske:2019cex}. Historically, after~first heavy ion DCE data were measured, a pair transfer scenario was favored, by which DCE reactions are assumed to proceed as a simultaneous mutual exchange of a proton pair in one direction and of a neutron pair in the other direction~\cite{Dasso:1986zza,Dasso:1985zvv}. The~transfer mechanism is a soft process driven by mean-field dynamics. The~minimal scenario is a sequence of two pair transfer reactions, e.g.,~$a(n,z)+A(N,Z)\to c(n-2,z)+C(N+2,Z)\to b(n-2,z+2)+B(N+2,Z-2)$, interfering with a second reaction path where the proton pair is exchanged first. Hence, in~leading order, the pair transfer scenario is at least of fourth order in the nucleon binding potentials. Single nucleon exchange processes are of even higher order. Transfer reaction mechanisms are most important in general at low incident energies where the kinematical conditions are favorable for probing mean-field dynamics. We will not consider further transfer DCEs which, in~fact, have been found to be negligible for the reactions considered here, as confirmed by recent experimental and theoretical investigations~\cite{Lay:2018dgv,Carbone:2020uxx}.

For a long time, pair transfer was thought to be the dominant heavy ion DCE reaction mechanism. After~the impressive successful use of heavy ion SCE reactions for spectroscopic studies (see, e.g.,~\cite{Lenske:2019cex}), a~first attempt to use heavy ion DCE reactions for spectroscopic purposes was made by Blomgren~et~al.~\cite{Blomgren:1995cux}, intending to measure the excitation of the spin-flip double Gamow--Teller resonance (DGTR). However, at~that time, the results were disappointing, which led the authors to rather pessimistic conclusions on the usefulness of heavy ion reaction for DCE studies. About a decade later, the~situation changed when the feasibility of DCE reactions and their potential for spectroscopic investigations was shown for the reaction $^{18}O+{}^{40}Ca \to ^{18}Ne+{}^{40}Ar$ at $T_{lab}=$~270~MeV by Cappuzzello~et~al.~\cite{Cappuzzello:2015ixp} in an experiment at LNS Catania. That experiment was important for narrowing down the conditions under which nuclear structure information can be extracted from data, as~is now the central goal of the NUMEN project~\cite{Cappuzzello:2018wek}.

\begin{figure}
\epsfig{file=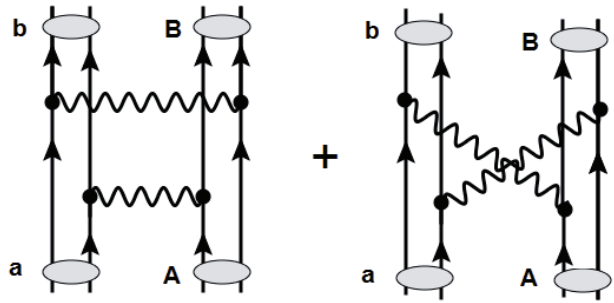, width=8.5cm}
\caption{Schematic graphical representation of a Double Single Charge Exchange (DSCE) reaction $a(N_a,Z_a)+A(N_A,Z_A)\to b(N_a\pm 2,Z_a\mp 2)+A(N_A\mp 2,Z_A\pm 2)$, proceeded by the sequential twofold action of the isovector NN T-matrix, indicated by wavy lines. Each of the interaction events acts similar to a one-body operator on the target and projectile, respectively. Note that the diagram on the right is related to left one by a change in time order. A~striking formal similarity to a $2\nu 2\beta$ nuclear matrix element (NME) is~apparent. }
\label{fig:DSCE}
\end{figure}

The observed angular distributions of Reference~\cite{Cappuzzello:2015ixp} in fact show a puzzling similarity to SCE reactions in shape and, to~a lesser extent,  in magnitude. Moreover, the~data cover a surprisingly large range of linear momentum transfer, extending up to about 500~MeV/c over the measured angular range. Thus, these properties demand a reaction mechanism different from low momentum-centered mean-field dynamics. The~appropriate candidate is charge exchange by hard collisional interactions, as provided by the mesonic DCE scenario, introduced for the first time in~\cite{Lenske:2018dkz}. Actually, as~pointed out in~\cite{Lenske:2018dkz,Lenske:2019cex}, there are two competing mesonic DCE reaction mechanisms. Here, we consider specifically the double single charge exchange (DSCE) scenario. As~illustrated in Figure~\ref{fig:DSCE}, the~DSCE reaction mechanism is given by two consecutive SCE events, both occurring half off-shell. As~was discussed in detail by Bellone~et~al.~\cite{Bellone:2020lal}, the DSCE process is of second order in the isovector NN T-matrix. The~measured angular distribution was described close to perfection in magnitude and very satisfactorily in shape by distorted wave theory, free space NN T-matrices, and~microscopic nuclear structure input. Hartree--Fock--Bogoliubov ground state densities were used for the optical potentials and response functions, and transition form factors were obtained by QRPA theory, as~discussed before in Reference~\cite{Lenske:2018jav}. Alternative approaches to the nuclear structure aspects of DCE reactions (as for DBD theory) are of course highly desirable. The~interacting boson model (IBM) belongs to the frequently used approaches in DBD theory. Using both eikonal and closure approximation, the aforementioned DCE reaction was analyzed in terms of an IBM–NME by Santopinto et al.~\cite{Santopinto:2018nyt}.

In~\cite{Bellone:2020lal}, the usual t-channel approach for calculating form factors was used, according to the scheme displayed in Figure~\ref{fig:DSCE}. That formulation is perfectly suited for the proper description of DSCE cross sections if the interest is focused on the reproduction or prediction of cross sections. However, that approach is not suitable for investigations and/or extraction of DBD nuclear matrix elements (NME) from cross sections. The~latter are connecting the two SCE-type vertices within the same nucleus, while the standard reaction theoretical approach is directed towards the description of the pair of vertices excited in the projectile and target in the first or the second steps of the DCE reaction. Thus, a~change from the conventional t-channel formulation to an appropriate s-channel formulation is required, not to the least as a necessary prerequisite for establishing the connection to the NME entering  DBD~theory.

Keeping this goal in mind, the~program of this paper is a purely theoretical one, namely to recast the second-order DSCE reaction amplitude into an s-channel representation. As~seen below, this requires a demanding amount of recoupling of various kinds of angular momenta, including the spectroscopic ones intrinsic to the involved nuclei and those describing the multipolarities acting on the relative motion degrees of freedom. Moreover, the~total number of form factors to be considered increases to the fourth power (or stronger) by the number of elementary NN-interaction vertices. Thus, the~full account of rank-0 central, rank-1 spin orbit, and rank-2 tensor NN-vertices results in general in a total of at least $4^4=256$ form factors, distributed half by half in the projectile and target. In~order to keep the presentation at a manageable level, we therefore restrict the discussion to the vertices of the rank-0 central interactions. As~was discussed already in References~\cite{Lenske:2018jav,Bellone:2020lal}, they involve already the complete set of relevant fundamental isovector vertices, describing non-spin flip $S=0$ Fermi-type and spin-flip $S=1$ Gamow--Teller-type nuclear transitions of any multipolarity. The~algebraic rules developed below can be used in the same manner also for more extended sets of NN-vertices. Practical calculations, e.g.,~those in Reference~\cite{Bellone:2020lal}, account of course for the complete set of~interactions.

The paper is organized as follows: The reaction theoretical aspects of DSCE reactions are presented in Section~\ref{sec:TheoDSCE}, adding additional theoretical background to our recent work~\cite{Bellone:2020lal}. The~DSCE reaction amplitude is derived and discussed in Section~\ref{sec:DSCE-NME}, leading to a factorized form that separates ion--ion interactions and nuclear matrix elements. A~key element of DSCE theory is discussed in Section~\ref{sec:Multipole}. The~reaction amplitude and the transition form factors are transformed from the $t$-channel to the $s$-channel representation, thus recasting the theory into a form compatible with and comparable to the formulations used in DBD theory. The~investigations lead also to the result that a rich spectrum of multipoles contributes to a DSCE reaction, confirming our previous numerical results on theoretical grounds~\cite{Bellone:2020lal}. The~physics content of the form factors and accordingly of DCE cross sections is investigated in Section~\ref{sec:Approx} by considering a few limiting cases. Form factors are derived in closure approximation. A~reduction scheme that  allows for a first-time derivation of DSCE multipole unit cross sections, which account for the reaction dynamical aspects and may serve in the future to extract DSCE--NME directly from data, is presented. A~few representative examples of unit cross sections are shown. The~work is summarized and an outlook is given in Section~\ref{sec:SumOut}. Certain coefficients resulting from the recoupling of angular momenta are found in Appendix \ref{app:AngCoup}.

\section{Theory of Sequential Double Charge Exchange~Reactions}\label{sec:TheoDSCE}

As depicted schematically in Figure~\ref{fig:DSCE}, the~double single charge exchange reactions are a sequence of two consecutive single charge exchange processes. After~the first SCE event, the system propagates in a combination of $\Delta Z=\pm 1$ configurations, concluded by a follow-up second charge exchange process. Each of the single charge exchange processes is induced by the two-body NN--isovector interaction $\mathcal{T}_{NN}$. The~T-matrix is used in a form given by one-body operators acting in the projectile and the target nucleus, respectively. For~a reaction $\alpha= a+A \to \beta=b+B$, the reaction amplitude is written down readily as a quantum mechanical second-order reaction matrix element~\cite{Bellone:2020lal}:
\be\label{eq:MDSCE}
\mathcal{M}^{(DSCE)}_{\alpha\beta}(\mathbf{k}_\alpha,\mathbf{k}_{\beta})=\lan \chi^{(-)}_\beta, bB|\mathcal{T}_{NN}\mathcal{G}^{(+)}_{aA}(\omega_\alpha)\mathcal{T}_{NN}|aA,\chi^{(+)}_{\alpha} \ran  .
\ee

Initial (ISI) and final state (FSI) interactions are taken into account by the distorted waves $\chi^{(\pm)}_{\alpha,\beta}$, depending on the center-of-mass (c.m.) momenta $\mathbf{k}_{\alpha,\beta}$ and obeying outgoing and incoming spherical wave boundary conditions, respectively. The~available c.m. energy is $\omega_\alpha=\sqrt{s_{aA}}$, where $s_{aA}=(T_{lab}+M_a+M_A)^2-p^2_{lab}$.

As discussed in~\cite{Lenske:2018jav}, we use an (anti-symmetrized) isovector NN T-matrix of the form
\be\label{eq:TNN}
\mathcal{T}_{NN}=\sum_{S=0,1,T=1}\bigg(\mathcal{T}_{ST}+\delta_{S1}\mathcal{T}_{Tn} \bigg)
\left(\tau^{(a)}_{+}\tau^{(A)}_{-}+ \tau^{(a)}_{-}\tau^{(A)}_{+}\right)
\ee

In non-relativistic notation, the~rank-0 central and rank-2 tensor interactions are 
 (In our notation, the~form factor of the rank-2 tensor interaction includes an additional factor $\sqrt{\frac{24\pi}{5}}$.)
\bea
\mathcal{T}_{ST}&=&V_{ST}\left[\bm{\sigma}_a\cdot\bm{\sigma}_A\right]^S\\
\mathcal{T}_{Tn}&=&V_{Tn}Y_2\cdot \left[\bm{\sigma}_a\otimes\bm{\sigma}_A\right]_2.
\eea

As indicated by the dot product, the~spherical harmonics $Y_{2M}$ has to be contracted with the rank-2 spin tensor to a total scalar. The~spin operators $\bm{\sigma}_{a,A}$ act in the projectile and the target.
Summation over all target and projectile nucleons is implicit. The~form factors $V_{ST}$ and $V_{Tn}$ are given by a superposition of meson-exchange propagators connecting projectile and target nucleons. The~strength factors are given by complex-valued coupling functionals in general depending on the energy in the NN center-of-momentum frame and the nuclear densities. Other operator structures such as two-body spin-orbit interactions will not be considered but are included~easily.

The intermediate propagator for the evolution of the intrinsic nuclear states and relative motion,
\be\label{eq:GaA}
\mathcal{G}^{(+)}_{aA}(\omega)=\frac{1}{\omega-\mathcal{H}_A-\mathcal{H}_a-\mathcal{H}_{aA}+i\eta},
\ee
is given by the nuclear Hamiltonians $\mathcal{H}_{a,A}$ and the relative motion Hamiltonian $\mathcal{H}_{aA}$. The~latter is described by an optical model Hamiltonian, $H_{opt}=T+U_{opt}$. With~the set of intermediate SCE-type states $\{|c\ran\}$ and $\{|C\ran\}$ in the projectile and target, respectively, we~obtain
\be
\mathcal{G}^{(+)}_{aA}(\omega)=\sum_{\gamma=\{c,C\}}|cC\ran G^{(+)}_{\gamma}(\omega)\lan cC|,
\ee
where the channel propagator is
\be\label{eq:GcC}
G^{(+)}_{\gamma}(\omega)=\frac{1}{\omega-E_c-E_C-H_{opt}+i\eta}.
\ee
$E_{c,C}=M_{c,C}+T_{c,C}$ is the total c.m. energies of the intermediate nuclei in states $c$ and $C$, respectively. As~discussed in~\cite{Bellone:2020lal}, by~means of the bi-orthogonal set of distorted waves $\{\tilde{\chi}^{(\pm)}_\gamma,\chi^{(\pm)}_\gamma\}$, the~reaction amplitude is finally obtained as
\be\label{eq:MDSCE1}
\mathcal{M}^{(2)}_{\alpha\beta}(\mathbf{k}_\alpha,\mathbf{k}_\beta)=\sum_{\gamma=\{c,C\}}
\int \frac{d^3k_\gamma}{(2\pi)^3}M^{(1)}_{\gamma\beta}(\mathbf{k}_\gamma,\mathbf{k}_\beta)
\frac{\tilde{S}^\dag_{\gamma}}{\omega_\alpha-E_c-E_C-T_\gamma+i\eta}
M^{(1)}_{\alpha\gamma}(\mathbf{k}_\alpha,\mathbf{k}_\gamma),
\ee
where $\tilde{S}^\dag_\gamma\sim \lan \tilde{\chi}^{(+)}_\gamma|\tilde{\chi}^{(-)}_\gamma\ran$ is the S-matrix element from the dual states $\tilde{\chi}^{(\pm)}_\gamma$, being solutions of $H^\dag_{opt}$. $T_\gamma$ denotes the kinetic energy related to the (off-shell) momentum $k_\gamma$. The half off-shell SCE amplitudes are of the form
\be
M^{(1)}_{\alpha\gamma}(\mathbf{k}_\gamma,\mathbf{k}_\alpha)=
\lan \chi^{(-)}_{\gamma}|F_{\alpha\gamma}|\chi^{(+)}_{\alpha}\ran,
\ee
with the transition form factor $F_{\alpha\gamma}=\lan cC|T_{NN}|aA\ran$.

The DSCE differential cross section (for unpolarized ions) is given as
\be\label{eq:dsigma}
d\sigma^{(DSCE)}_{\alpha\beta}=\frac{m_\alpha m_\beta}{(2\pi\hbar^2)^2}\frac{k_\beta}{k_\alpha}\frac{1}{(2J_a+1)(2J_A+1)}
\sum_{M_a,M_A\in \alpha;M_b,M_B\in \beta}{\left|M^{(2)}_{\alpha\beta}(\mathbf{k}_\alpha,\mathbf{k}_\beta)\right|^2}d\Omega,
\ee
averaged over the initial nuclear spin states ($J_{a,A}$ and $M_{a,A}$) and summed over the final nuclear spin states  ($J_{b,B} and M_{b,B}$, respectively). Reduced masses in the incident and exit channel, respectively, are denoted by $m_{\alpha,\beta}$.

\section{The Heavy Ion DSCE Reaction~Amplitude}\label{sec:DSCE-NME}

In momentum space, the~operator structure of the isovector part of the NN T-matrix is determined in all tensorial parts by the operators~\cite{Lenske:2018jav}
\be\label{eq:RST_FDE}
R_{ST}(\mathbf{p})=e^{i\mathbf{p}\cdot \mathbf{r}}\bm{\sigma}^S\tau^\pm .
\ee

With the nuclear transition form factors
$F^{(DE)}_{ST}(\mathbf{p})=\lan E|R_{ST}(\mathbf{p})|D \ran $,
the half off-shell SCE amplitudes become
\bea\label{eq:M1ab}
&&M^{(1)}_{\alpha\gamma}(\mathbf{k}_\gamma,\mathbf{k}_\alpha)=
\int d^3p  D_{\alpha\gamma}(\mathbf{p})\sum_{S=0,1,T=1}\\
&&
\left(V_{ST}(p^2)F^{(ac)}_{ST}(\mathbf{p})\cdot F^{(AC)}_{ST}(\mathbf{p})+
\delta_{S1}V_{Tn}(p^2)Y_2(\hat{\mathbf{p}})\cdot\left[F^{(ac)}_{ST}(\mathbf{p})\otimes F^{(AC)}_{ST}(\mathbf{p})\right]_2  \right),\nonumber
\eea
and accordingly for $M^{(1)}_{\gamma\beta}$. The~above---on first sight, unusual---form was chosen in virtue of displaying the factorization of the SCE reaction amplitude into a distortion coefficient $D_{\alpha\gamma}$, containing the elastic ion--ion interactions and---in brackets---the nuclear transition form factors, describing intranuclear SCE dynamics. The~full information on elastic ion--ion interactions is contained in the distortion coefficients:
\be\label{eq:Dab}
D_{\alpha\gamma}(\mathbf{p})=\frac{1}{(2\pi)^3}\lan\chi^{(-)}_{\gamma}|e^{i\mathbf{\mathbf{p}}\cdot \mathbf{r}_\alpha}|\chi^{(+)}_{\alpha}\ran .
\ee

They can be considered an extension of the S-matrix concept into the off-shell~region. 

By contour integration, the propagator Equation~\ref{eq:GaA} is separated into the intrinsic nuclear and the relative motion propagators,
\be
\mathcal{G}^{(+)}_{aA}=\oint_{C^{+}}\frac{d\nu}{2i\pi}G^{(+)}_{opt}(\omega_\alpha-\nu)G_{aA}(\nu).
\ee
by which a formal separation of the relative motion and intrinsic nuclear evolution is achieved. The~integration path $C_+$ extends over the upper half of the complex $\nu$--plane. Applying the momentum representation, Equation~\ref{eq:M1ab}, the~DSCE reaction amplitude becomes
\bea\label{eq:MDSCE_sep}
&&M^{(2)}_{\beta\alpha}(\mathbf{k}_\beta,\mathbf{k}_\alpha)=
\int d^3p_1d^3p_2 \oint_{C_+}\frac{d\nu}{2i\pi}\sum_{S_1,S_2}\Pi^{(S_2S_1)}_{\alpha\beta}(\mathbf{p}_2,\mathbf{p}_1;\nu)\\ &&\int\frac{d^3k_\gamma}{(2\pi)^3}D_{\beta\gamma}(\mathbf{p}_2)V_{S_2T}(p^2_2)\frac{\tilde{S}^\dag_\gamma}{\omega_\alpha-\nu-T_\gamma+i\eta}
 D_{\gamma\alpha}(\mathbf{p}_1)V_{S_1T}(p^2_1)
.\nonumber
\eea
where contributions of  higher rank tensor operators have been left out for the reasons discussed in the Introduction. In~numerical calculations, the full spectrum of tensor operators is, of~course, taken into~account.

The projectile and target sequential SCE responses are now contained in the nuclear polarization tensor:
\bea\label{eq:NucTensor}
\Pi^{(S_2S_1)}_{\alpha\beta}(\mathbf{p}_2,\mathbf{p}_1;\nu)=
\sum_{cC}
\frac{F^{(BC)}_{S_2}(\mathbf{p}_2)\cdot F^{(bc)}_{S_2}(\mathbf{p}_2)F^{(ca)}_{S_1}(\mathbf{p}_1)\cdot
F^{(CA)}_{S_1}(\mathbf{p}_1)}{\nu-(E_A-E_C+E_a-E_c)},
\eea
combining, however, both projectile and target transitions.
As indicated by the dot products, a total spin-scalar tensor is~obtained.

\section{Multipole Structure of the Transition Form Factors and Nuclear Matrix~Elements}\label{sec:Multipole}

The result of Equation~\ref{eq:NucTensor} is in fact perfectly well suited for DCE reaction calculations, as in~\cite{Bellone:2020lal}. The~focus of this section is to clarify the relation of a DSCE reaction to nuclear matrix elements (NMEs) of the projectile and target. Hence, we develop a formalism by which the contributions of the two nuclei to the combined ion--ion NME can be separated. Such a program requires decomposing and rearranging the nuclear tensor of Equation~\ref{eq:NucTensor} in an appropriate manner.
The momentum representation provides the suitable~formalism.

As seen by the results of the previous section, the~polarization tensor is given by products of Fourier--Bessel transforms of transition densities. The~momentum structure of the transition densities is probed by the operators $R_{ST}$ (Equation~\ref{eq:RST_FDE}). By~expanding the plane waves into multipoles, we find the spin-scalar ($S=0$) Fermi-like and spin-vector ($S=1$) Gamow--Teller-like isovector ($T=1$) one-body operators:
\be\label{eq:TLSJ}
T_{(\ell S)IN}(\mathbf{r},p)=\sum_{m_\ell M}\left[i^\ell j_\ell(pr) Y_{\ell}(\mathbf{\hat{r}})\otimes\bm{\sigma}^S\right]_{IN}\bm{\tau},
\ee
where $j_\ell(x)$ denotes a spherical Bessel function of order $\ell$. The~nuclear SCE form factors~become \vspace{6pt}
\be
F^{(DE)}_{ST}(\mathbf{p})=4\pi\sum_{\ell m IN}Y^*_{\ell m}(\hat{\mathbf{p}})
\left(\ell m S \mu |I N \right)\left(J_E M_E J_D -M_D |I N \right)(-1)^{J_D-M_D}R^{J_DJ_E}_{\ell S I}(p)
\ee
where the Wigner--Eckardt theorem was used to derive the reduced matrix elements:
\be\label{eq:RJDJE}
R^{J_DJ_E}_{\ell S I}(p)=\frac{1}{\widehat{I}}\lan J_E||T_{(\ell S)I}||J_D\ran.
\ee

We use the notation $\widehat{I}=\sqrt{2I+1}$.

In order to clarify the physics content, we  emphasize that the matrix elements \mbox{(Equation~\ref{eq:RJDJE})} are in fact momentum-dependent transition form factors. As~such, they do not follow the rules known from beta-decay on the enhancement or suppression of multipolarities already by the operator structure alone. Only for $p\to 0$ the transition operators of Equation~\ref{eq:TLSJ} approach the long--wave length limit underlying the weak and the electromagnetic operators commonly used in nuclear structure theory. For~sufficiently large $p$---as easily realized in a heavy ion reaction---essentially all multipole operators are of the same magnitude as dictated by the asymptotics of the Bessel--Riccati~functions.

The standard ordering of transitions in a DSCE reaction follows the scheme indicated in Figure~\ref{fig:DSCE}, as~imposed by meson exchange. However, in~order to comply with the goal to access projectile and target DSCE nuclear matrix elements, a~regrouping and recoupling of terms and correspondingly of angular momenta is required in order to follow the evolution of the intrinsic nuclear states instead of focusing on meson exchange. In~other words, a~change in representation from the $t$-channel to the $s$-channel has to be~performed.

The complexities of the second-order process are reflected of course in a correspondingly involved formalism. As~a rule of thumb, momenta and quantities such as operators and quantum numbers of spins and angular momenta related to the first and the second SCE vertexes will be denoted by indices $1$ and $2$, respectively. Quantities related to processes in the projectile or target nuclei are usually denoted by the index $a$ and $A$, respectively, occasionally complemented by indices $c,b$ and $C,B$ if states in the projectile-like and the target-like intermediate and final nuclei have to be~distinguished.

Considering only the central spin--scalar and spin--vector interactions by the reasons discussed in the Introduction, the~result is
\be
\sum^{S_1+S_2}_{S=|S_1-S_2|,M}(-)^{S_1+S_2-S+M}
\left[F^{(BC)}_{(S_2T)}(\mathbf{p}_2)\otimes F^{(CA)}_{(S_1T)}(\mathbf{p}_1)\right]_{SM}
\left[F^{(bc)}_{(S_2T)}(\mathbf{p}_2)\otimes F^{(ca)}_{(S_1T)}(\mathbf{p}_1)\right]_{S-M},
\ee
thus now being in an order appropriate for the separation of projectile and target response functions. A~second contour integration is used to separate completely projectile and target NMEs:
\begingroup\makeatletter\def\f@size{9.5}\check@mathfonts
\def\maketag@@@#1{\hbox{\m@th\normalsize\normalfont#1}}%
\bea\label{eq:Pi_recoup}
&&\Pi^{S_1S_2}_{\alpha\beta}(\mathbf{p}_2,\mathbf{p}_1;\nu)=\\
&&\sum_{SM_S}(-)^{S_1+S_2-S}\oint_{C^+}\frac{d\omega}{2i\pi}
(-)^{M_S}\Pi^{(AB)}_{(S_1S_2)SM_S}(\mathbf{p}_2,\mathbf{p}_1;\omega)\cdot \Pi^{(ab)}_{(S_1S_2)S-M_S}(\mathbf{p}_2,\mathbf{p}_1;\nu -\omega)\nonumber,
\eea
\endgroup
where care has been taken in maintaining a total spin-scalar result. The~target tensor, for~example, is
\be\label{eq:PiAB}
\Pi^{(AB)}_{(S_1S_2)SM}(\mathbf{p}_2,\mathbf{p}_1;\omega)=\sum_C
\frac{\left[F^{(BC)}_{S_2}(\mathbf{p}_2)\otimes F^{(CA)}_{S_1}(\mathbf{p}_1)\right]_{SM}}{\omega-(E_A-E_C)}.
\ee
and the projectile tensor is defined accordingly.
Replacing $E_A-E_C\sim M_A-M_C$ and re-interpreting $\omega$ as the lepton energy, a striking similarity to the NME of $2\nu 2\beta$ decay, e.g.,~\cite{Tomoda:1990rs}, is immediately~identified.

By several steps of angular momentum recoupling, the nominator of Equation~\ref{eq:PiAB} is finally obtained as a superposition of irreducible multipole components:
\bea
&&\left[F^{(BC)}_{S_2}(\mathbf{p}_2)\otimes F^{(CA)}_{S_1}(\mathbf{p}_1)\right]_{SM}=\\
&&(-)^{J_A-M_A}
\sum_{I_AN_A,LM_L}(-)^{I_A-N_A}\left(J_AM_AJ_BM_B|I_AN_A\right)\left(L M_L S M_S|I_A N_A \right)
 \nonumber\\
&&\times\sum_{\ell_1\ell_2}
(-)^{M_L} \mathcal{Y}_{(\ell_1\ell_2)LM_L}(\hat{\mathbf{p}}_1,\hat{\mathbf{p}}_2)
R^{J_AJ_CJ_BI_A}_{(S_1S_2)S;(\ell_1\ell_2)L}(p_1,p_2)
\nonumber
\eea
where we introduced the bi-spherical harmonics:
\be
\mathcal{Y}_{(\ell_1\ell_2)LM}(\hat{\mathbf{p}}_1,\hat{\mathbf{p}}_2)=
\left[Y_{\ell_1}(\hat{\mathbf{p}}_1)\otimes Y_{\ell_2}(\hat{\mathbf{p}}_2)  \right]_{LM}.
\ee

The reduced form factors themselves are defined by a superposition of multipole contributions:
\bea
R^{J_AJ_CJ_BI_A}_{(S_1S_2)S;(\ell_1\ell_2)L}(p_1,p_2)=
(4\pi)^2\sum_{I_1I_2}Z^{J_AJ_CJ_B}_{LSI_A}(\ell_1\ell_2;S_1S_2;I_1I_2)
R^{(J_BJ_C)}_{\ell_2S_2I_2}(p_2)R^{(J_CJ_A)}_{\ell_1S_1I_1}(p_1),
\eea
resulting from the coupling of transferred spin and orbital angular momenta to the total angular momentum transfers $I_{1,2}$ in the first and second SCE interactions, respectively. The~recoupling coefficients are defined in Appendix \ref{app:AngCoup}.

We introduce the multipole polarization propagators
\be
\Pi^{J_AJ_BI_A}_{S_1S_2S,LM_L}(\mathbf{p}_1,\mathbf{p}_2,\omega)=\sum_{\ell_1\ell_2}
\mathcal{Y}_{(\ell_1\ell_2)LM}(\hat{\mathbf{p}}_1,\hat{\mathbf{p}}_2)
\overline{\Pi}^{J_AJ_BI_A}_{S_1S_2S,\ell_1\ell_2L_A}(p_1,p_2,\omega)
\ee
with the reduced polarization propagators
\be\label{eq:PiAB_reduced}
\overline{\Pi}^{J_AJ_BI_A}_{S_1S_2S,\ell_1\ell_2L_A}(p_1,p_2,\omega)=\sum_{C,J_C}
\frac{R^{J_AJ_CJ_BI_A}_{(S_1S_2)S;(\ell_1\ell_2)L}(p_1,p_2)}{\omega-(E_A-E_C)}
\ee
by which Equation~\ref{eq:PiAB} becomes
\begingroup\makeatletter\def\f@size{9.5}\check@mathfonts
\def\maketag@@@#1{\hbox{\m@th\normalsize\normalfont#1}}%
\bea\label{eq:PiAB_MultiFF}
&&\Pi^{(AB)}_{(S_1S_2)SM_S}(\mathbf{p}_2,\mathbf{p}_1;\omega)=\\
&&\sum_{I_AN_A,LM_{L}}(-)^{I_A-N_A}\left(J_AM_AJ_BM_B|I_AN_A\right)\left(L M_{L} S M_S|I_A N_A \right)
\Pi^{J_AJ_BI_A}_{S_1S_2S,LM_L}(\mathbf{p}_1,\mathbf{p}_2,\omega)\nonumber
\eea
\endgroup

Inserting  Equation~\ref{eq:PiAB_MultiFF} into Equation~\ref{eq:Pi_recoup} and the corresponding expressions for the complementary sequence $a\to c \to b$, the~summation over the magnetic spin quantum numbers $M_S$ can be performed (see Appendix \ref{app:AngCoup}) and we obtain the intermediate result
\bea\label{eq:PI_S1S2}
&&\Pi^{(S_1S_2)}_{\alpha\beta}(\mathbf{p}_2,\mathbf{p}_1;\nu)=\\
&&\sum_{I_AN_A,I_aN_A}(-)^{I_A-N_A}(-)^{I_a-N_a}
\left(J_AM_AJ_BM_B|I_AN_A  \right)\left(J_aM_aJ_bM_b|I_aN_a  \right)\nonumber\\
&&\times \sum_{S}(-)^{S_1+S_2-S}\sum_{L_A,L_a,\lambda\mu}
U^{I_AI_aS}_{(L_AL_a)\lambda}\left(L_AM_{L_A}L_aM_{L_a}|\lambda\mu  \right)\left(I_AM_{A}I_aM_{a}|\lambda\mu  \right)\nonumber\\
&&\times\oint_{C^+}\frac{d\omega}{2i\pi}
\left[\Pi^{J_AJ_BI_A}_{S_1S_2S,L_A}(\mathbf{p}_1,\mathbf{p}_2,\omega)\otimes \Pi^{J_aJ_bI_a}_{S_1S_2S,L_a}(\mathbf{p}_1,\mathbf{p}_2,\omega-\nu)   \right]_{\lambda\mu}
\nonumber
\eea

The coupling indicated in the last line of the above formula is finally exploited to recouple the two bi-spherical harmonics into a single one, as~shown in Appendix~\ref{app:AngCoup}. As~a consequence, the~angular dependencies are stripped off the nuclear tensors, and~we find
\bea
&&\Pi^{(S_1S_2)}_{\alpha\beta}(\mathbf{p}_2,\mathbf{p}_1;\nu)=\sum_{I_AN_A,I_aN_A,\lambda\mu}(-)^{I_A-N_A}(-)^{I_a-N_a}\\
&&\left(J_AM_AJ_BM_B|I_AN_A  \right)\left(J_aM_aJ_bM_b|I_aN_a  \right)
\left(I_AN_{A}I_aN_{a}|\lambda\mu  \right)
\mathcal{F}^{J_AJ_BI_A,J_aJ_bI_a}_{S_1S_2;\lambda\mu}(\mathbf{p}_2,\mathbf{p}_1;\nu), \nonumber
\eea
with the transition form factor of total multipolarity $\lambda$
\bea\label{eq:Falfabeta}
&&\mathcal{F}^{J_AJ_BI_A,J_aJ_bI_a}_{S_1S_2;\lambda\mu}(\mathbf{p}_2,\mathbf{p}_1;\nu)=\\
&&\sum_{L_{13}L_{24}}
\mathcal{Y}_{(L_{13}L_{24})}(\hat{\mathbf{p}}_1,\hat{\mathbf{p}}_2)\sum_{S,L_AL_a}\sum_{\ell_1\ell_3,\ell_2\ell_4}
A^{I_AI_aS}_{L_{13}L_{24}\lambda}(\ell_1\ell_3,\ell_2\ell_4,L_AL_a)\nonumber\\
&&\times \oint\frac{d\omega}{2i\pi}\overline{\Pi}^{J_AJ_BI_A}_{S_1S_2S,\ell_1\ell_2L_A}(p_1,p_2,\omega)
\overline{\Pi}^{J_aJ_bI_a}_{S_1S_2S,\ell_3\ell_4L_a}(p_1,p_2,\omega-\nu).\nonumber
\eea
where $A^{I_AI_aS}_{L_{13}L_{24}\lambda}(\ell_1\ell_3,\ell_2\ell_4,L_AL_a)$ is found in Appendix~ \ref{app:AngCoup}.
This allows us to define the reduced reaction amplitudes
\bea\label{eq_MDSCE_RED}
&&\mathcal{M}^{J_AJ_BI_A}_{J_aJ_bI_a;\lambda\mu}(\mathbf{k}_\alpha,\mathbf{k}_\beta)=
\int d^3p_1d^3p_2
\oint_{C_+}\frac{d\nu}{2i\pi}\sum_{S_1,S_2}\mathcal{F}^{J_AJ_BI_A,J_aJ_bI_a}_{S_1S_2;\lambda\mu}(\mathbf{p}_2,\mathbf{p}_1;\nu) \\ &&\times \int\frac{d^3k_\gamma}{(2\pi)^3}D_{\beta\gamma}(\mathbf{p}_2)V_{S_2T}(p^2_2)\frac{\tilde{S}^\dag_\gamma}{\omega_\alpha-\nu-T_\gamma+i\eta}
D_{\gamma\alpha}(\mathbf{p}_1)V_{S_1T}(p^2_1).\nonumber
\eea

By comparison to Equation~\ref{eq:NucTensor}, the~essence of the exercise is that we have achieved a reduction in the $a+A\to b+B$ nuclear transition tensor to a form displaying explicitly the target and projectile response functions and their multipole structure. In~addition, the~$A\to B$ and $a\to b$ angular momentum coupling coefficients have been split off such that, for the angular distribution (Equation~\ref{eq:dsigma}), the~summations over the magnetic quantum numbers can be performed, resulting in
\be\label{eq:dsigmaRED}
d\sigma^{(DSCE)}_{\alpha\beta}=\frac{m_\alpha m_\beta} {(2\pi\hbar^2)^2} \frac{k_\beta}{k_\alpha}\frac{1}{(2J_a+1)(2J_A+1)}
\sum_{I_a,I_A;\lambda\mu}{\left|
\mathcal{M}^{J_AJ_BI_A}_{J_aJ_bI_a;\lambda\mu}(\mathbf{k}_\alpha,\mathbf{k}_\beta)\right|^2}d\Omega.
\ee

As a closing remark to this section, we emphasize that the formulation has been kept very general, intending to cover for future use the full multipole spectrum. For~special cases, especially for $J^\pi_{A,a}=0^+ \to J^\pi_{B,b}=0^+$ transitions, the~situation simplifies considerably. The~total angular momentum transfer is constrained to $I_A=0$, which for the total orbital and spin angular momentum transfer implies the two combinations $L=0,S=0$ and $L=2,S=2$, respectively. In~the first case, the~intermediate channels are restricted to sequential excitations of Fermi modes or Gamow--Teller modes where $\ell_1=\ell_2$ and \mbox{$\ell_{1,2}=I_{1,2}$}, i.e.,~the same multipolarity is excited in each of the two SCE steps. The \mbox{$L=2,S=2$} case is accessible only by sequential Gamow--Teller-type transitions of the same total multipolarity, $I^\pi_1=I^\pi_2$.
For a $0^+\to 0^+$ reaction, the combination $L=1,S=1$ is forbidden by parity~conservation.

\subsection{Nuclear Structure~Aspects}\label{ssec:NucMod}

In order to evaluate the polarization tensors, nuclear wave functions are required for the involved states. Nuclear ground states are described by Hartree--Fock--Bogoliubov (HFB) theory, as discussed in~\cite{Hofmann:1997zu,Tsoneva:2017kaj,Lenske:2019ubp}. The~SCE excited  states are obtained by QRPA calculations; see, e.g.,~\cite{Lenske:2018jav,Bellone:2020lal} for recent~results.

The DCE parent and daughter nuclei are connected by an isospin rotation perpendicular to the $I_3$-axis. The~rotation is such that the isospin in each nucleus is changed by two units but the total isospin as defined by the incident projectile--target system is conserved, of~course. Since isospin is a conserved symmetry in strong interactions, we are eligible to expect that the ground states of the parent and the daughter nuclei are related in leading order by a rather transformation, changing by a rotation in quasiparticle space,  e.g.,~a pair of protons into a pair of neutrons (or vice~versa). Hence, in~the HFB mean-field picture, the final state reached by a $\Delta Z=-2$  transition is dominantly given by an $n^{2}p^{-2}$-configuration in the valence shells. Thus, assuming for $|J_AM_A\ran=|0\ran$ a $0^+$ ground state, states in the $\Delta Z=-2$ daughter nucleus will be considered as  a 4-quasi particle configuration with respect to $|0\ran$.
Thus, we use
\be
|J_BM_B\ran=N_B[\alpha^+_{j_{n_1}}\alpha^+_{j_{n_2}}\alpha^+_{j_{p_1}}\alpha^+_{j_{p_2}}]_{J_BM_B}|0\ran
\ee
where $\alpha^+_{jm}$ is a single quasiparticle operator. The~proper normalization is taken care of by the constant $N_B$. Typically, such two particle--two hole states are rather stable against perturbations. Thus, good approximation admixtures of higher-order configurations, caused by residual interactions inducing  core polarization, can be~neglected.

In DBD theory, the~ground state of the nucleus $B$ is usually treated as the quasiparticle vacuum state of the daughter nucleus. However, in~a DCE reaction, that point of view does not match the sequential character of the transition. We emphasize that a DSCE reaction probes the 4~quasiparticle (QP) content of the states in $B$ with respect to the parent nucleus. A~clear advantage of that picture is that the whole spectrum of final states is accessible by the same theoretical~methods.

The $|\Delta Z|=1$ intermediate states are of a more complex structure. Starting from an even--even ground state---as is common practice---the odd--odd character of the SCE states and collectivity have to be taken into account. Thus, residual interactions have to be included into the theoretical description for which QRPA theory is an appropriate approach. Hence, the~intermediate states $|k,J_CM_C\ran=\Omega^\dag_{kJ_CM_C}|0\ran$ of energy $E_{kJ_C}$ are described by acting with a two-quasiparticle (2QP) QRPA operator $\Omega^\dag_{kJ_CM_C}$ onto the ground state:
\bea
\Omega^\dag_{k,J_CM_C}=\sum_{j_pj_n}x^{J_C}_\gamma(j_pj_n) Q^\dag_{J_CM_C}(j_pj_n)-y^{J_C*}_\gamma(j_pj_n) \widetilde{Q}_{J_CM_C}(j_pj_n)
\eea
with the 2QP operators $Q^\dag_{J_CM_C}(j_pj_n)=\left[\alpha^\dag_{j_p}\otimes \alpha^\dag_{j_n} \right]_{J_CM_C}$ and the time-reversed state $\widetilde{Q}_{JM}=(-)^{J+M}Q_{J-M}$. The~4QP states in $B$ are grouped and coupled accordingly:
\be
|J_BM_B\ran=\sum_{J_1J_2}C^{(J_1J_2)J_B}_{j_{p_1}j_{n_1}j_{p_2}j_{n_2}}
\left[Q^\dag_{J_1}(j_{p_1}j_{n_1})\otimes Q^\dag_{J_2}(j_{p_2}j_{n_2})\right]_{J_BM_B}|0\ran,
\ee
where the coefficient $C$ accounts for recoupling and normalization.
Hence, even in the simplest case of a 4QP configuration given by $0^+$ pairs of protons and neutrons, a~rich spectrum of multipolarities is encountered when transformed to the particle--hole~representation.

In second quantization, the~transition operators (Equation~\ref{eq:TLSJ}) become in the $pn^{-1}$-channel
\be\label{eq:TLSJ2nd}
T^{(pn)}_{(LS)JM} =\sum_{j_pj_n}R^{j_pj_n}_{LSJ}(p)\left(u_{j_p}v_{j_n}Q^\dag_{JM}(j_pj_n)+u_{j_n}v_{j_p}\widetilde{Q}_{JM}(j_pj_n)  \right),
\ee
while in the $np^{-1}$-channel, we find
\be\label{eq:TLSJ2nd}
T^{(np)}_{(LS)JM} =\sum_{j_pj_n}R^{j_nj_p}_{LSJ}(p)\left(u_{j_n}v_{j_p}Q^\dag_{JM}(j_pj_n)+u_{j_p}v_{j_n}\widetilde{Q}_{JM}(j_pj_n)  \right).
\ee
where scattering terms $\sim \alpha^+_q\alpha_{q'}$ have been neglected.
We use the same angular momentum coupling scheme for the $\tau_+$ and the $\tau_{-}$ cases and exploit
the reduced matrix elements obeying the relation $R^{j_nj_p}_{LSJ}(p)=(-)^{S}R^{j_pj_n*}_{LSJ}(p)$. The~Bogoliubov--Valatin QP 
amplitudes are denoted by $u_j$ and $v_j$, respectively. Within~the 2QP--representation, the~evaluation of the two sequential SCEs transition is a comparatively easy task, especially for a $0^+$ reference state. For~$A\to C$ transitions of $pn^{-1}$ character, the~NME of Equation~\ref{eq:RJDJE} is
\be
R^{J_AJ_C}_{LSJ}(p)=\sum_{j_pj_n}R^{j_pj_n}_{LSJ}(p)\left(x^{J_C*}_{j_pj_n}u_{j_p}v_{j_n}+(-)^{S}y^{J_C}_{j_pj_n}u_{j_n}v_{j_p}  \right).
\ee

Since the $C\to B$ transitions start from an already excited nucleus, the~form factors are of a quite different structure: the form factors of the second $pn^{-1}$ transition are superpositions of contributions given by one 2QP NME times an overlap amplitude of the second 2QP pair with the previous SCE excitation.
\bea
&&R^{J_CJ_B}_{LSJ}(p)=\sum_{J_1J_2}C^{(J_1J_2)J_B}_{j_{p_1}j_{n_1}j_{p_2}j_{n_2}}\\
&&\times\left(u_{j_{p_1}}v_{j_{n_1}}R^{j_{p_1}j_{n_1}}_{LSJ_1}(p)x^{J_C}_{j_{p_2}j_{n_2}}\delta_{J_2J_C}+
u_{j_{p_2}}v_{j_{n_2}}R^{j_{p_2}j_{n_2}}_{LSJ_2}(p)x^{J_C}_{j_{p_1}j_{n_1}}\delta_{J_1J_C}  \right).\nonumber
\eea

The quasiparticle rescattering contributions, neglected here, would lead to form factors involving a quasiparticle from the intermediate $J_C$ phonon and a quasiparticle from the final $J_B$ configuration. Different from the $ph$-type form factors, the~scattering terms are of the order $\mathcal{O}(u_pu_n)$ and $\mathcal{O}(v_pv_n)$, respectively. Thus, these transitions proceed by decomposing the state vectors of the intermediate configurations into their single quasiparticle components, thereby destroying the coherence of the~transition.

\subsection{Brief on Spectral Properties of DCE~Transitions}\label{ssec:SpecProp}

As an important message from the above results, we notice that the structure of the final $B$-configurations plays an essential role in selecting the admissible intermediate SCE states. This, of~course, affects also the reaction mechanism because the structure of the final DCE state $B$ determines the path through the pool of intermediate SCE states by constraining the accessible~multipolarities.

As an example, we consider more closely the
DCE reaction ${}^{18}O+{}^{40}Ca\to {}^{18}Ne+{}^{40}Ar$, which was observed a few years ago~\cite{Cappuzzello:2015ixp} and studied theoretically recently in~\cite{Bellone:2020lal}. The~incident channel involves only ($2s,1d$)-shell nuclei. In~the exit channel, the ($2p,1f$)-nucleus $^{40}Ar$  and the ($2s,1d$)--hell ejectile ${}^{18}Ne$ are present. The HFB results predict that ${}^{40}Ar(0^+,g.s.)$ is given with respect to ${}^{40}Ca$ in good approximation by two hole states in the $1d_{3/2}$-proton shell and two particle states in the $1f_{7/2}$-neutron shell. Hence, the~recoupling leads to 2QP proton--neutron states of negative parity, implying a clear preference for negative parity intermediate states in ${}^{40}K$. On~the projectile side, ${}^{18}Ne(0^+,g.s.)$ may be considered in leading order as a $(1d_{5/2}(p))^{2}(1d_{5/2}(n))^{-2}$ relative to ${}^{18}O(0^+,g.s.)$, as predicted by our HFB calculation. Since only positive parity 2QP-states are involved, this implies a selectivity for a route through positive parity states in ${}^{18}F$  (Spectral distributions for ${}^{40}K$ and ${}^{18}F$ are found in~\cite{Lenske:2018jav}).

\section{Approximations}\label{sec:Approx}

The matrix elements derived in the previous sections are of a rather demanding mathematical (and numerical) structure. Appropriately chosen approximations are of great help to identify the leading physical quantities and to understand the essential features of such an involved second-order reaction. The~purpose of this section is to exemplify a few interesting aspects of sequential DCE reactions by discussing approximations exploiting the fact that the intermediate states in the projectile and target are essential but~remain unresolved, serving merely as a kind of pool of background states, resembling to some extent a heat bath. Hence, their influence on the reaction amplitude may be treated in an averaged~manner.

\subsection{Nuclear NME in Closure~Approximation}\label{ssec:Closure}
The multipole polarization tensors (Equation~\ref{eq:PiAB_reduced}) may be manipulated in a meaningful manner by
introducing a yet to be determined mean excitation energy $\overline{\omega}_{C}=\lan E_A- E_C\ran$. A~power series expansion  results in
\be
\overline{\Pi}^{J_AJ_BI_A}_{S_1S_2S,\ell_1\ell_2L_A}(p_1,p_2,\omega)=\frac{1}{\omega -\overline{\omega}_{C}}
\sum_{C,J_C}R^{J_AJ_CJ_BI_A}_{(S_1S_2)S;(\ell_1\ell_2)L}(p_1,p_2)
\left(1-\frac{\omega_C-\overline{\omega}_{C}}{\omega-\overline{\omega}_{C}}\ldots
\right).
\ee

By the first term, we recover the closure approximation, namely the unconstrained summation over the full set of intermediate states and multipolarities $\{C,J_C\}$:
\be
\overline{R}^{J_AJ_BI_A}_{S_1S_2S,\ell_1\ell_2L_A}(p_1,p_2)=
\sum_{C,J_C}R^{J_AJ_CJ_BI_A}_{(S_1S_2)S;(\ell_1\ell_2)L}(p_1,p_2).
\ee

In principle, $\overline{\omega}_C$ can be derived from the spectral distribution of intermediate states. Cancellation of the second term is achieved by choosing as reference energy
\be
\overline{\omega}_{C}=\frac{\sum_{C,J_C}\omega_{C}R^{J_AJ_CJ_BI_A}_{(S_1S_2)S;(\ell_1\ell_2)L}(p_1,p_2)}
{\sum_{C,J_C}R^{J_AJ_CJ_BI_A}_{(S_1S_2)S;(\ell_1\ell_2)L}(p_1,p_2)}
\ee
which will also minimize the contributions of the higher-order terms. Dependencies on the momenta $p_{1,2}$ will be canceled in leading order. Thus, $\overline{\omega}_{C}$ is fixed by the ratio of the energy-weighted and the non-energy-weighted sum rules of the full spectrum of intermediate states (to a good approximation, $\overline{\omega}_{C}$ can be derived from the (observed) SCE spectrum of the intermediate $Z\pm 1$-nuclei, provided that the range of measured excitation energy is sufficiently large). Applying the same procedure also to the second ion, the~product of the leading order terms leads to an energy denominator $\sim$$1/(\omega-\overline{\omega}_{C})(\nu -\omega-\overline{\omega}_{c})$. The~contour integral in
Equation~\ref{eq_MDSCE_RED} can be performed and leads to
\bea\label{eq:MDSCE_Clos}
&&\mathcal{M}^{J_AJ_BI_A}_{J_aJ_bI_a}(\mathbf{k}_\alpha,\mathbf{k}_\beta)=
\int d^3p_1d^3p_2
\sum_{S_1,S_2}\mathcal{R}^{J_AJ_BI_A,J_aJ_bI_a}_{S_1S_2}(\mathbf{p}_2,\mathbf{p}_1) \\ &&\times \int\frac{d^3k_\gamma}{(2\pi)^3}N_{\beta\gamma}(\mathbf{p}_2)V_{S_2T}(p^2_2)
\frac{\tilde{S}^\dag_\gamma}{\omega_\alpha-\overline{\omega}_\gamma-T_\gamma+i\eta}
N_{\gamma\alpha}(\mathbf{p}_1)V_{S_1T}(p^2_1).\nonumber
\eea
where $\overline{\omega}_\gamma=\overline{\omega}_C+\overline{\omega}_c$. In~closure approximation, the form factor (Equation~\ref{eq:Falfabeta}) transforms into
\bea\label{eq:Ralfabeta}
&&\mathcal{R}^{J_AJ_BI_A,J_aJ_bI_a}_{S_1S_2;\lambda\mu}(\mathbf{p}_2,\mathbf{p}_1)=\\
&&\sum_{L_{13}L_{24}}
\mathcal{Y}_{(L_{13}L_{24})\lambda\mu}(\hat{\mathbf{p}}_1,\hat{\mathbf{p}}_2)
\sum_{S,L_AL_a}\sum_{\ell_1\ell_3,\ell_2\ell_4}A^{I_AI_aS}_{L_{13}L_{24}\lambda}(\ell_1\ell_3,\ell_2\ell_4,L_AL_a) \nonumber\\
&&\times \overline{R}^{J_AJ_BI_A}_{S_1S_2S,\ell_1\ell_2L_A}(p_1,p_2)
\overline{R}^{J_aJ_bI_a}_{S_1S_2S,\ell_3\ell_4L_a}(p_1,p_2).\nonumber
\eea

\subsection{Effective Form~Factors}
An important difference between hadronic and leptonic reactions is the quite different momentum structure and strength of interactions. As~found in~\cite{Bellone:2020lal}, a large number of states of high angular momenta are excited in heavy ion DCE reactions while beta decay processes are dominated by low multipolarities, rarely larger than $J=2$. Another important property of SCE and DCE reactions is the excitation of high-lying states in the continuum region. Hence, this quasi-statistical nature of the spectrum of intermediate states induces self--averaging effects by which characteristics of individual transitions will be largely washed out. These effects may be exploited to further reduce the complex structure of the reaction amplitude. A~meaningful approach is to approximate the transition form factors by average multipole form factors $D^{(N)}_{\ell}(p)$ separately for each nucleus $N=A,a$. Thus, as~the simplest possible approach, we demand equality of the state-dependent density form factors and a mean density form factor at an appropriate momentum transfer $p=p_\ell$
\be
R^{J_DJ_E}_{\ell SI}(p_\ell)=N^{J_DJ_E}_{\ell SI}D^{(N)}_{\ell}(p_\ell)
\ee
by which the state and multipole-dependent amplitude $N^{J_DJ_E}_{\ell SI}$ is defined. The~choice of the matching momentum $p_L$ is uncritical as long as the above relation is used and the effective density form factor $D^{(N)}_{\ell }$ accounts  realistically for the essential features of the momentum structure. A~case of practical relevance is the choice $p_\ell \to 0$ at which the Bessel--Riccati functions approach the limit $j_\ell (pr)p^{-\ell }\to r^\ell /(2\ell +1)!!$. Thus, by~this choice, the case of the long-wave length limit of weak (and electromagnetic) multipole operators is used as a reference~point.

That kind of parametrization leads to decoupling of the state dependence, now contained in the amplitudes $N^{J_DJ_E}_{\ell SI}$, from~the momentum dependence, now described by the effective form factors $D^{(N)}_{\ell}(p)$. The~reduced polarization propagators (Equation~\ref{eq:PiAB_reduced}) emerge as bilinear forms of the effective form factors, and the multipole propagators become
\be\label{eq:PiAB_DD}
\Pi^{J_AJ_BI_A}_{S_1S_2S,LM_L}(\mathbf{p}_1,\mathbf{p}_2,\omega)=\sum_{\ell_1\ell_2}
\left[\mathcal{D}^{(A)}_{\ell_1}(\mathbf{p}_1)\otimes \mathcal{D}^{(A)}_{\ell_2}(\mathbf{p}_2)\right]_{LM}
\overline{\Pi}^{J_AJ_BI_A}_{S_1S_2S,\ell_1\ell_2L_A}(p_{\ell_1},p_{\ell_2},\omega)
\ee
with $\mathcal{D}^{(N)}_{\ell m}(\mathbf{p})=D^{(N)}_\ell(p)Y_{\ell m}(\hat{\mathbf{p}})$. The~transition form factors (Equation~\ref{eq:Falfabeta}) are changed to
\bea\label{eq:Falfabeta_DD}
&&\mathcal{F}^{J_AJ_BI_A,J_aJ_bI_a}_{S_1S_2,\lambda\mu}(\mathbf{p}_2,\mathbf{p}_1;\nu)\approx \\
&&\sum_{\ell_1\ell_3,\ell_2\ell_4}\sum_{L_{13}L_{24}}
\left[\mathcal{D}^{\ell_1\ell_3L_{13}}_{\alpha\gamma}(\mathbf{p}_1)\otimes
\mathcal{D}^{\ell_2\ell_4L_{24}}_{\gamma\beta}(\mathbf{p}_2)\right]_{\lambda\mu}
\sum_{SL_AL_a}
A^{I_AI_aS}_{L_{13}L_{24}\lambda}(\ell_1\ell_3,\ell_2\ell_4,L_AL_a)\nonumber\\
&&\times \oint\frac{d\omega}{2i\pi}\overline{\Pi}^{J_AJ_BI_A}_{S_1S_2S,\ell_1\ell_2L_A}(p_{\ell_1},p_{\ell_2},\omega)
\overline{\Pi}^{J_aJ_bI_a}_{S_1S_2S,\ell_3\ell_4L_a}(p_{\ell_1},p_{\ell_2},\omega-\nu).\nonumber
\eea

The products of the projectile and target form factors have been rearranged to 
\be
\mathcal{D}^{\ell_1\ell_3L_{13}M_{13}}_{\alpha\gamma}(\mathbf{p}_1)=
D^{(A)}_{\ell_1}(p_1)D^{(a)}_{\ell_3}(p_1)Y_{L_{13}M_{13}}(\hat{\mathbf{p}}_1),
\ee 
and $\mathcal{D}^{\ell_2\ell_4L_{24}}_{\gamma\beta}(\mathbf{p}_2)$ is defined~accordingly.

A highly interesting result is found by combining the effective form factor method and the closure approach. With~that combination, we obtain a full separation of reaction and nuclear dynamics (although still being coupled on the level of angular momenta). Integrals over $d^3p_{1,2}$ can be performed and restore the second-order DW reaction amplitude but now describes the scattering on the effective form factors, e.g.,~for the first interaction
\be
\overline{\mathcal{F}}^{\ell_1\ell_3L_{13}M_{13}}_{ST}(\mathbf{r}_\alpha)=\int \frac{d^3p}{(2\pi)^3}
e^{-i\mathbf{p}\cdot \mathbf{r}_\alpha}
V_{ST}(p^2)\mathcal{D}^{\ell_1\ell_3L_{13}M_{13}}_{\alpha\gamma}(\mathbf{p}).
\ee

\begin{figure}
\includegraphics[width=8.5cm]{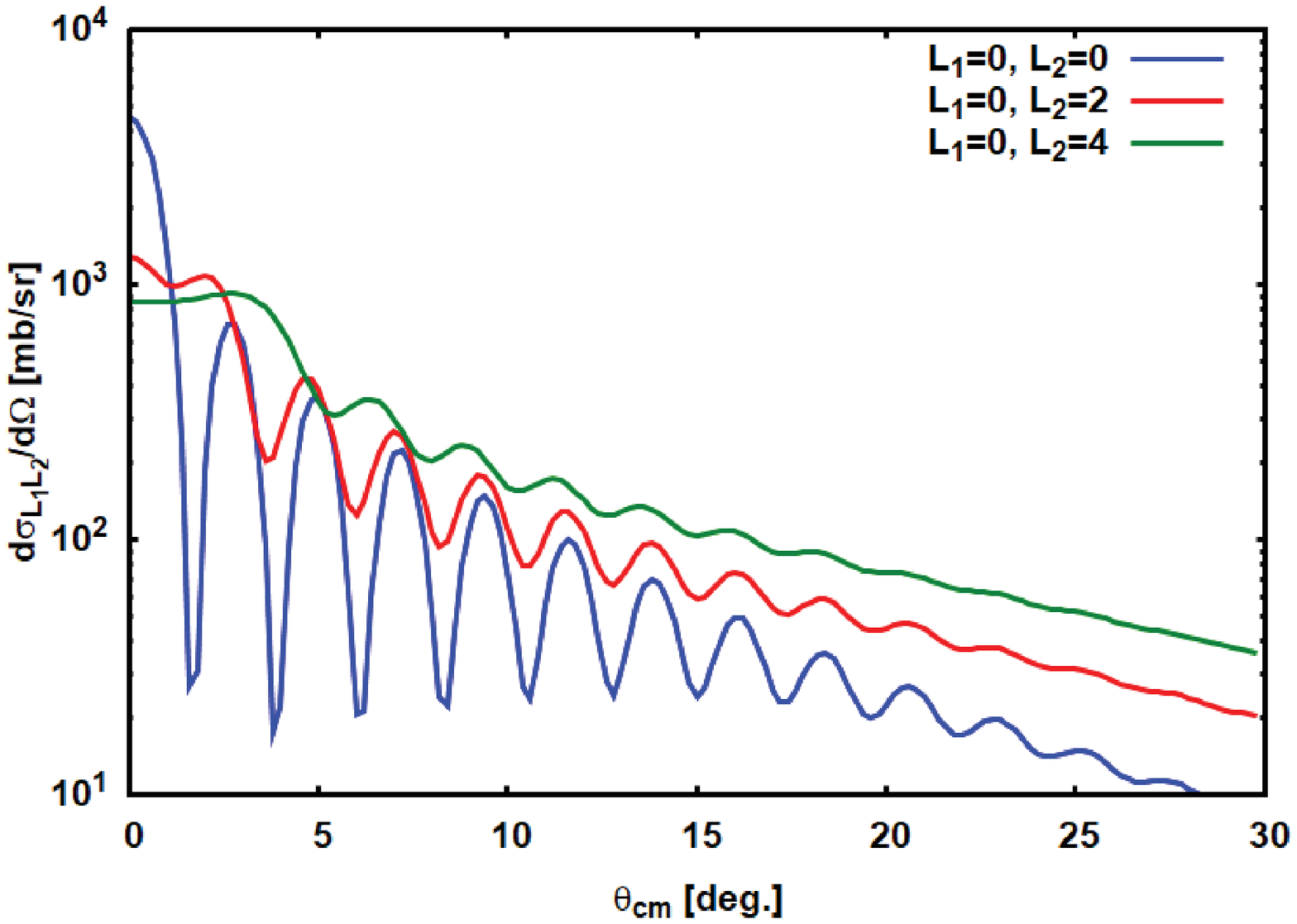}\\
\includegraphics[width=8.5cm]{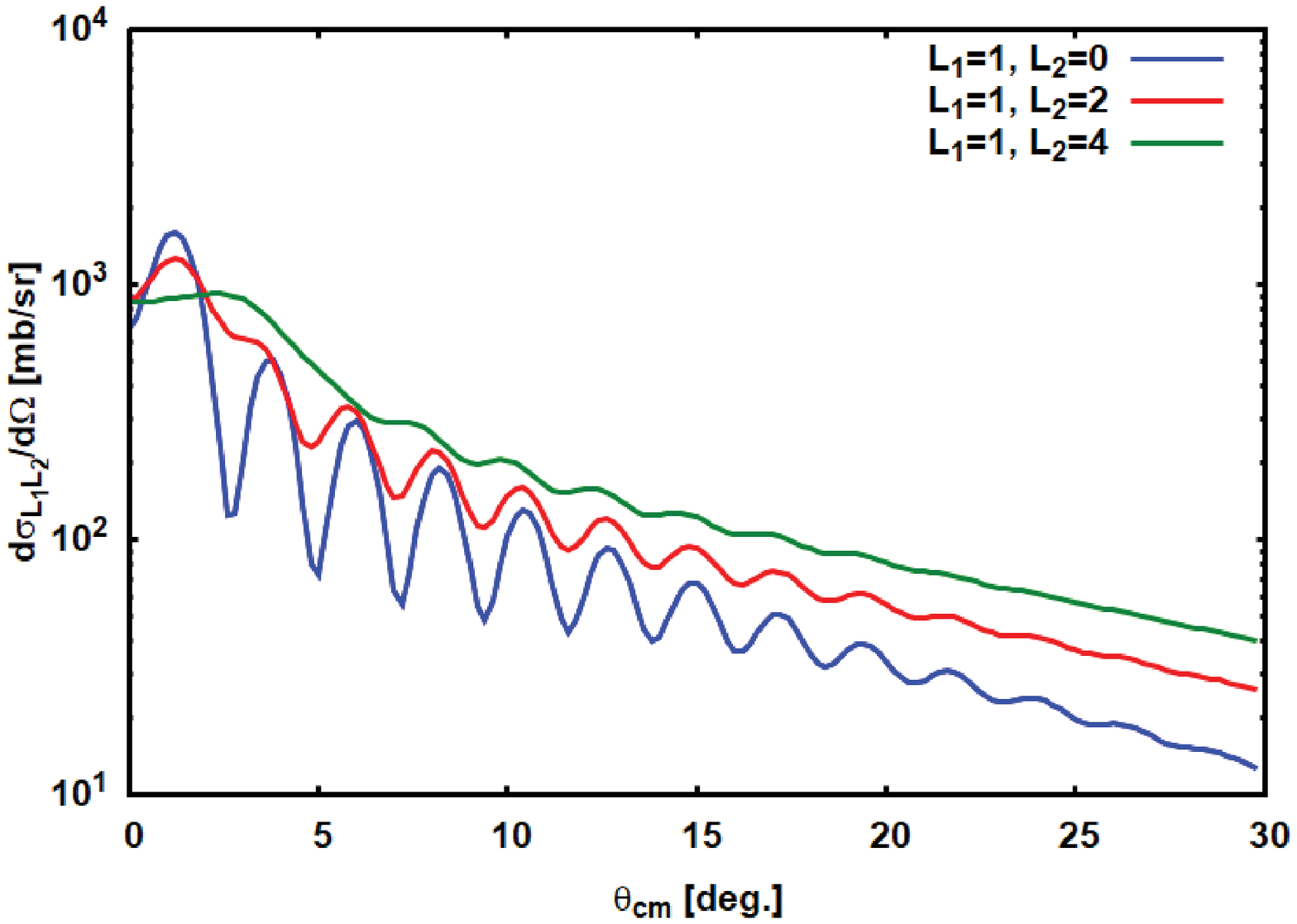}\\
\includegraphics[width=8.5cm]{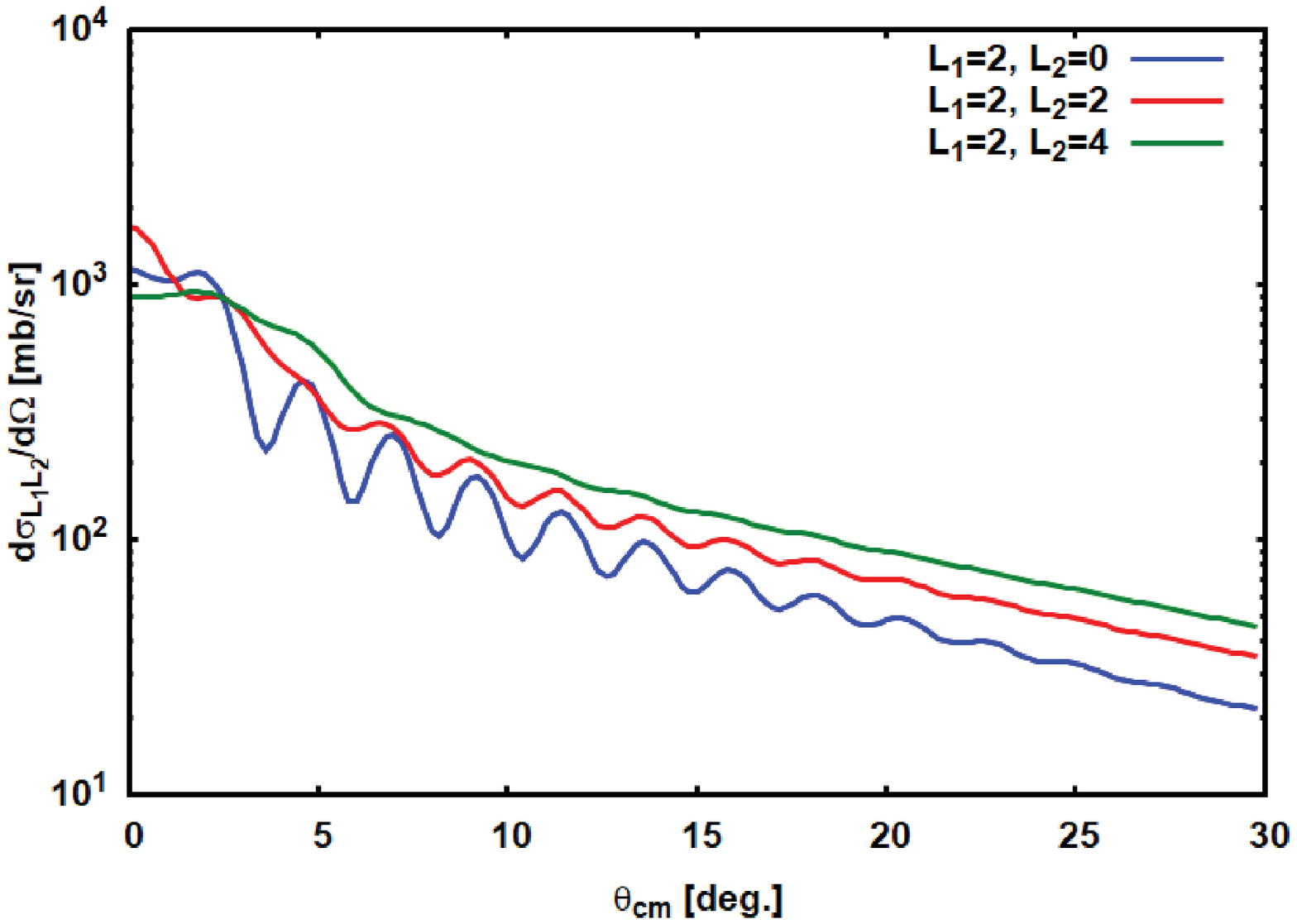}\\
\caption{Second-order DSCE unit strength cross sections for the reaction
$^{18}O+{}^{40}Ca\to {}^{18}N+{}^{40}Ar$ at $T_{lab}=270$~MeV. From~top to bottom, results  are shown for total angular momentum transfer in the first single charge exchange (SCE) interaction $L_1=0,2$ and the second SCE interaction $L_2=0,2,4$, respectively. The~average excitation energy was chosen as $\overline{\omega}_\gamma=10$~MeV. The~angular range corresponds to momentum transfers up to 1000~MeV/c. Optical potentials and transition potentials are calculated in a double folding approach by using the (newly derived) nucleon--nucleon (NN) T-matrix at $T_{lab}=15$~MeV, parameterized as in References~\cite{Love:1981gb,Love:1987zx}. Optical potentials are calculated with Hartree--Fock--Bogoliubov (HFB) ground state densities according to Reference~\cite{Lenske:2018jav}. The~cross sections are calculated by using average form factors derived from QRPA transition densities as discussed in the text. For~comparison, SCE unit strength cross sections are shown in Figure \ref{fig:dsigSCE}.}
\label{fig:dsigLL}
\end{figure}

Correspondingly, the second step form factor $\overline{\mathcal{F}}^{\ell_2\ell_4L_{24}M_{24}}_{ST}(\mathbf{r}_\beta)$ is obtained. The~expansion of the intermediate channel propagator is reversed and
Equation~\ref{eq:MDSCE_Clos} becomes
\bea\label{eq:MDSCE_Clos_DD}
&&\mathcal{M}^{J_AJ_BI_A}_{J_aJ_bI_a;\lambda\mu}(\mathbf{k}_\alpha,\mathbf{k}_\beta)\approx\\
&&\sum_{S_1,S_2}\sum_{\ell_1\ell_3,\ell_2\ell_4}\sum_{L_{13}L_{24}}
\lan \chi^{(-)}_\beta | \left[\overline{\mathcal{F}}^{\ell_2\ell_4L_{24}}_{S_2T}G_{opt}(\omega_\alpha-\overline{\omega}_\gamma)
\otimes\overline{\mathcal{F}}^{\ell_1\ell_3L_{13}}_{S_1T}\right]_{\lambda \mu}|\chi^{(+)}_\alpha\ran\nonumber\\
&&\times S^{I_AI_a,S_1S_2}_{J_AJ_B,J_aJ_b}(\ell_1\ell_3,\ell_2\ell_4,L_{13}L_{24}\lambda)\nonumber
\eea
with the spectroscopic DSCE amplitude
\begingroup\makeatletter\def\f@size{9.5}\check@mathfonts
\def\maketag@@@#1{\hbox{\m@th\normalsize\normalfont#1}}%
\bea
&&S^{I_AI_a,S_1S_2}_{J_AJ_B,J_aJ_b}(\ell_1\ell_3,\ell_2\ell_4,L_{13}L_{24}\lambda)=\\
&&\sum_{S,L_AL_a}\sum_{\ell_1\ell_3,\ell_2\ell_4}A^{I_AI_aS}_{L_{13}L_{24}\lambda}(\ell_1\ell_3,\ell_2\ell_4,L_AL_a)
\overline{R}^{J_AJ_BI_A}_{S_1S_2S,\ell_1\ell_2L_A}(p_{\ell_1},p_{\ell_2})
\overline{R}^{J_aJ_bI_a}_{S_1S_2S,\ell_3\ell_4L_a}(p_{\ell_1},p_{\ell_2}).\nonumber
\eea
\endgroup

As mentioned before, these expression simplify considerably for special combinations of nuclear states, among~which reactions starting from $J^\pi=0^+$ ground states are of particular interest. However, in~order to explore the wealth of DCE data to be expected for the near future, the~whole spectrum of final states, at least in the target, has to be understood. In~any case, the~spin-scalar and the spin-vector channels have to be taken into account, leading in general to a coherent superposition of spin-dependent form~factors.

In Equation~\ref{eq:MDSCE_Clos_DD}, the~reaction amplitudes
\be
\overline{\mathcal{M}}^{\ell_2\ell_4 L_{24}S_2}_{\ell_1\ell_3L_{13}S_1;\lambda\mu}(\mathbf{k}_\alpha,\mathbf{k}_\beta)=
\lan \chi^{(-)}_\beta | \left[\overline{\mathcal{F}}^{\ell_2\ell_4L_{24}}_{S_2T}G_{opt}(\omega_\alpha-\overline{\omega}_\gamma)
\otimes\overline{\mathcal{F}}^{\ell_1\ell_3L_{13}}_{S_1T}\right]_{\lambda \mu}|\chi^{(+)}_\alpha\ran
\ee
are within our formalism the \emph{unit strength amplitudes} of second-order DW theory. In~those cases where interference terms can be neglected, they lead to \emph{unit strength cross sections}:
\be\label{eq:dsigmaUnit}
d\sigma^{\ell_2\ell_4 L_{24}S_2}_{\ell_1\ell_3 L_{13}S_1;\lambda}=
\frac{m_\alpha m_\beta} {(2\pi\hbar^2)^2} \frac{k_\beta}{k_\alpha}\frac{1}{(2J_a+1)(2J_A+1)}
\sum_{\mu}{\left|\overline{\mathcal{M}}^{\ell_2\ell_4 L_{24}S_2}_{\ell_1\ell_3L_{13}S_1;\lambda\mu}(\mathbf{k}_\alpha,\mathbf{k}_\beta)\right|^2}d\Omega,
\ee

\begin{figure}
\includegraphics[width=8cm]{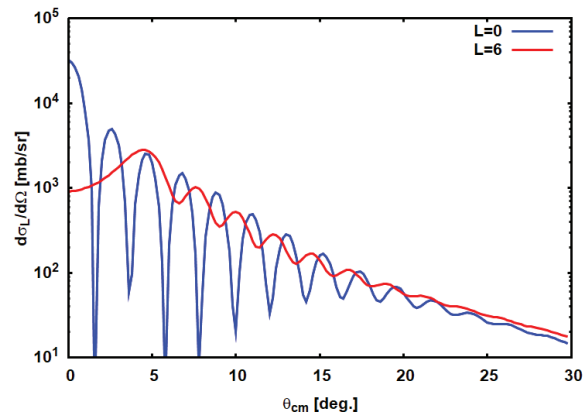}
\caption{First-order SCE unit strength cross sections for the reaction
$^{18}O+{}^{40}Ca\to {}^{18}N+{}^{40}Ar$ at $T_{lab}=270$~MeV. The~angular range corresponds to momentum transfers up to 1000~MeV/c. Optical potentials and transition potentials are calculated in a double folding approach by using the (newly derived) nucleon--nucleon (NN) T-matrix at $T_{lab}=15$~MeV, parameterized as in References~\cite{Love:1981gb,Love:1987zx}. Optical potentials are calculated with Hartree--Fock--Bogoliubov (HFB) ground state densities according to Reference~\cite{Lenske:2018jav}. The~cross sections are calculated by using average form factors derived from QRPA transition densities as discussed in the text.}
\label{fig:dsigSCE}
\end{figure}

Representative results of DSCE unit strength differential cross sections are shown in \mbox{Figure~\ref{fig:dsigLL}} for the reaction $^{18}O+{}^{40}Ca\to {}^{18}N+{}^{40}Ar$ at $T_{lab}=270$~MeV. For comparison, SCE unit cross sections are depicted in \mbox{Figure~\ref{fig:dsigSCE}} The~magnitudes are almost independent of the ($L_{13},L_{24}$) combinations of first- and second-step total angular momentum transfers, while the shapes are strongly affected by the~multipolarities.

In future studies, the~unit cross sections (Equation~\ref{eq:dsigmaUnit}) may be used to extract information on nuclear form factors directly from data. However, there are a number of caveats to keep in mind. The~neglection of interference effects will lead to systematic errors, which could be estimated from the quality of description of the angular distributions: if interference effects are important, they will be reflected in the diffraction structures. Since the unit cross sections are defined for specific multipolarities, their use in an empirically analysis requires energy distributions of high resolution, e.g.,~available for light ion SCE reactions. Moreover, a~clear multipole decomposition of spectra requires measuring spectral distributions at several scattering angles. In~other words, double differential cross sections need to be measured over a sufficiently large range of scattering angles and a large range of excitation energies. The~angular range will be decisive for  access to the momentum structures of form factors, characterizing their multipole structure. A~broad energy range is needed to explore the spectral distributions of the multipolarities. Another point to remember is that the DCE response is always a combined response of the target and projectile. For~SCE reactions, that complication is well under control, as reviewed in~\cite{Lenske:2019cex}. For~DSCE reactions, that problem is easy to handle for theory but corresponding experimental techniques have to be~developed.

\section{Summary}\label{sec:SumOut}
For the first time, a~consistent theoretical description of heavy ion sequential double charge exchange reactions was presented. The~theory is focused on collisional DCE reactions mediated by a sequence of two consecutive charge-transforming SCE events due to the exchange of isovector mesons. Reaction dynamics is described by second-order distorted wave theory. The~main focus was on a consistent microscopic formulation of reaction and intrinsic nuclear dynamics. A~scheme was introduced for the separation of target and projectile NMEs, which was achieved by a recoupling from the $t$-channel to an $s$-channel formulation, presented here for the first time. A~general scheme was used to describe the nuclear transition form factors. Aspects of nuclear DSCE spectroscopy were discussed by using description based on quasiparticle mean-field and QRPA theory. Essential features of the form factors were investigated theoretically. The~properties of the DSCE transition form factors were investigated in detail by exploring several limiting cases. Unit strength DSCE cross sections were derived, which under neglection of interference effects may serve to extract nuclear matrix elements directly from data by a multipole decomposition of spectral~distributions.

The theoretical methods are of general character allowing us to describe transitions of arbitrary combinations of multipolarities in the projectile and target. It is worth emphasizing that the theory presented here is constrained  neither to a specific projectile--target combination nor to specific regions of incident energies. The~reaction theoretical parts do not rely on a specific kind of nuclear structure model but is open for input of any kind of structure model. That option will be exploited in future work, for~example, to~compare systematically DSCE results for diver approaches to nuclear matrix elements and transition for factors, thus encircling the systematic uncertainties related to the choice of structure models. As~a concrete project, a comparison of QRPA and IBM transition form factors is in preparation. The~dependence of DSCE results on optical potentials is another important topic to be explored further. Thus, theoretical methods are at hand, ready to describe DSCE data becoming available in the near~future.

\appendix
\section{Angular Momentum~Couplings}\label{app:AngCoup}
The multipole decomposition of the nuclear transition form factors is
\bea
&&R^{J_AJ_CJ_BI_A}_{\ell_1\ell_2L;S_1S_2S}(p_1,p_2)=\\
&&\sum_{I_1I_2}(-)^{I_1+I_2} \widehat{I}_1\widehat{I}_2 W(J_AI_1J_BI_2;J_CI_A)
\left\{
    \begin{array}{ccc}
      \ell_1 & S_1 & I_1 \\
      \ell_2 & S_2 & I_2 \\
      L &  S & I_A \\
    \end{array}
  \right\}
R^{(J_BJ_C)}_{\ell_2S_2I_2}(p_2)R^{(J_CJ_A)}_{\ell_1S_1I_1}(p_1),\nonumber
\eea
including a Racah-W and a 9-j symbol. By
\be
Z^{J_AJ_CJ_B}_{LSI_A}(\ell_1\ell_2;S_1S_2;I_1I_2)=(-)^{I_1+I_2} \widehat{I}_1\widehat{I}_2 W(J_AI_1J_BI_2;J_CI_A)
\left\{
    \begin{array}{ccc}
      \ell_1 & S_1 & I_1 \\
      \ell_2 & S_2 & I_2 \\
      L &  S & I_A \\
    \end{array}
  \right\}
\ee
we obtain
\bea
R^{J_AJ_CJ_BI_A}_{\ell_1\ell_2L;S_1S_2S}(p_1,p_2)=\sum_{I_1I_2}Z^{J_AJ_CJ_B}_{LSI_A}(\ell_1\ell_2;S_1S_2;I_1I_2)
R^{(J_BJ_C)}_{\ell_2S_2I_2}(p_2)R^{(J_CJ_A)}_{\ell_1S_1I_1}(p_1).
\eea

The summation over the spin-magnetic quantum numbers $M_S$, indicated in \mbox{Equation~\ref{eq:PiAB_MultiFF}}, leads to
\bea
&&\sum_{M_S}(-)^{M_S}\left(L_A M_{L_A} S M_S|I_A N_A \right)\left(L_a M_{L_a} S -M_S|I_a N_a \right)=\\
&&\sum_{\lambda\mu}U^{I_AI_aS}_{(L_AL_a)\lambda}\left(L_AM_{L_A}L_aM_{L_a}|\lambda\mu  \right)\left(I_AN_{A}I_aM_{a}|\lambda\mu  \right),\nonumber
\eea
thus combining the intranuclear angular momentum transfers $L_{A,a}$ to the total orbital angular momentum transfer $\lambda$, as~expressed by a Clebsch--Gordan coefficient, where
\be
U^{I_AI_aS}_{(L_aL_a)\lambda}=(-)^{L_a+I_A-\lambda}\widehat{I}_A\widehat{I}_aW(L_AI_AL_aI_a;S\lambda)
\ee

Finally, the~above result is used to couple the product of bi-spherical harmonics to a single total angular momentum transfer $\lambda$, resulting in
\bea
&&\sum_{M_{L_A}M_{L_a}}\left(L_AM_{L_A}L_aM_{L_a}|\lambda\mu  \right)
\mathcal{Y}_{(\ell_1\ell_2)L_AM_{L_A}}(\mathbf{p}_1,\mathbf{p}_2) \otimes
\mathcal{Y}_{(\ell_3\ell_4)L_aM_{L_a}}(\mathbf{p}_1,\mathbf{p}_2)=\\
&&\sum_{L_{13}L_{24}}X_{L_{13}L_{24}\lambda}(\ell_1\ell_3,\ell_2\ell_4,L_AL_a)
\mathcal{Y}_{(L_{13}L_{24})\lambda \mu}(\mathbf{p}_1,\mathbf{p}_2),
\eea
where a recoupling coefficient is obtained
\be
X_{L_{13}L_{24}\lambda}(\ell_1\ell_3,\ell_2\ell_4,L_AL_a)=A_{\ell_1\ell_3L_{13}}A_{\ell_2\ell_4L_{24}}\widehat{L}_{13}\widehat{L}_{24}\widehat{L}_{A}\widehat{L}_{a}
\left\{
    \begin{array}{ccc}
      \ell_1 & \ell_3 & L_{13} \\
      \ell_2 & \ell_4 & L_{24} \\
       L_A & L_a & \lambda \\
    \end{array}
  \right\}
\ee
and
\be
A_{\ell_i\ell_j L_{ij}}=\frac{\widehat{\ell}_i\widehat{\ell}_j}{\sqrt{4\pi}\widehat{L}_{ij}}
\left(\ell_i 0 \ell_j 0|L_{ij} 0 \right).
\ee

Finally, we define the coupling coefficients
\bea
A^{I_AI_aS}_{L_{13}L_{24}\lambda}(\ell_1\ell_3,\ell_2\ell_4,L_AL_a)=
U^{I_AI_aS}_{(L_AL_a)\lambda}X_{L_{13}L_{24}\lambda}(\ell_1\ell_3,\ell_2\ell_4,L_AL_a).
\eea

\acknowledgments{
This research was funded in~part by DFG, contract Le 439/16, and~Alexander-von-Humboldt~Foundation, and INFN.

Inspiring discussions with F. Cappuzello, M. Cavallaro, and~C. Agodi are acknowledged. H.L. is grateful for the hospitality at LNS Catania and for the support by INFN. M. C. acknowledges the support from the European
Unions Horizon 2020 research and innovation
programme under Grant Agreement No. 654002.
 }

\end{document}